\documentclass{AIMS_LENCOS}
\usepackage{amsmath}
  \usepackage{paralist}
  \usepackage{graphics} 
  \usepackage{epsfig} 
 \usepackage[colorlinks=true]{hyperref}
\hypersetup{urlcolor=blue, citecolor=red}

\def\currentvolume{X}
 \def\currentissue{X}
  \def\currentyear{200X}
   \def\currentmonth{XX}
    \def\ppages{X--XX}

\def\R{\mathbb{R}}
\def\L{\mathcal{L}}

\def\pa{\partial}

\def\da{ \frac{\mathrm{d} \alpha}{\mathrm{d}t}}
\def\db{ \frac{\mathrm{d} \beta}{\mathrm{d}t}}
\def\dc{ \frac{\mathrm{d} \chi}{\mathrm{d}t}}
\def\de{ \frac{\mathrm{d} \eta}{\mathrm{d}t}}
\def\dk{ \frac{\mathrm{d} k}{\mathrm{d}t}}

\def\pc{ \frac{\partial}{\partial \chi} }
\def\pe{ \frac{\partial}{\partial \eta} }

\def\rootfig{./eps/}

\theoremstyle{definition}

\title[Variational approximations of bifurcations]
       {Variational approximations of bifurcations of asymmetric solitons in cubic--quintic nonlinear Schr\"odinger lattices}
       
\author[C. Chong  and  D.E. Pelinovsky ]{}


 \email{christopher.chong@math.uni-karlsruhe.de}
 \email{dmpeli@math.mcmaster.ca}

\thanks{The work of C.C. was partially supported by the Humboldt Research Foundation
and the Graduiertenkolleg 1294: Analysis, simulation and design of nano-technological processes,
which is sponsored by the Deutsche Forschungsgemeinschaft (DFG) and the Land Baden-W\"urttemberg.
The work of D.P. was partially supported by the NSERC grant.}

\begin{document}
\maketitle

\centerline{\scshape C. Chong}
\medskip
{\footnotesize
 \centerline{Fakult\"at f\"ur Mathematik, Universit\"at Karlsruhe }
   \centerline{ Karlsruhe 76128, Germany}
} 

\medskip

\centerline{\scshape D.E. Pelinovsky}
\medskip
{\footnotesize
 \centerline{Department of Mathematics, McMaster University}
   \centerline{Hamilton, Ont., Canada L8S 4K1}
}

\bigskip


\begin{abstract}
Using a variational approximation we study discrete solitons of a
nonlinear Schr\"odinger lattice with a cubic-quintic nonlinearity.
Using an ansatz with six parameters we are able to approximate
bifurcations of asymmetric solutions connecting site-centered and
bond-centered solutions and resulting in the exchange of their
stability. We show that the numerically exact and variational
approximations are quite close for solitons of small powers.
\end{abstract}

\section{Introduction}
\label{intro}

The variational approximation (VA) has long been used as a
semi-analytic technique to approximate solitary wave solutions of
nonlinear evolution equations with an underlying Hamiltonian
structure \cite{Ma02}. There have been a number of papers
exploring the VA with four parameters as a relevant approximation
of localized modes in discrete nonlinear Schr\"odinger
(DNLS) equations \cite{JKFM,MM91,PKMF03}. Kaup \cite{Ka05}
extended the variational approximation with six parameters that
allowed him to construct not only site-centered solutions (also
called on-site solitons) from \cite{MM91} but also the
bond-centered solutions (solitons centered at a midpoint between
two adjacent sites also known as inter-site solitons).

Site-centered and bond-centered solitons were recently considered
in the context of the DNLS equations with competing cubic focusing and
quintic defocusing nonlinearities both in the space of one
\cite{Ricardo} and two \cite{Ch09} lattice dimensions. It was
found that the two branches exchange their stability while
continued with respect to the underlying parameters.
A salient feature of this stability exchange is that the
two branches of site-centered and bond-centered solitions do not
intersect directly but are connected by an intermediate branch of
asymmetric solitons. It was argued in \cite{Ch09} that the
discrete solitons have enhanced mobility near the regimes of
stability inversion. These properties were originally discovered in the
DNLS equations with a saturable nonlinearity both in the space of one
\cite{Stepic} and two \cite{VJ} dimensions as well as for the DNLS
equations with on-site and next-site cubic nonlinearities in the space of
one dimension \cite{OJE}. It was recently shown for the same model
in the space of two dimensions \cite{OJ} that stability inversion
and asymmetric solutions may not lead to enhanced mobility of
discrete solitons if the bifurcation points of the solution
branches are widely separated in the parameter space.

It is the purpose of this work to apply Kaup's variational method
with six parameters from \cite{Ka05} to explain bifurcations of
asymmetric solutions and stability exchange of site-centered and
bond-centered discrete solitons in the context of the
one-dimensional cubic--quintic DNLS equation. In this sense, our work is a
complement to the previous paper \cite{Ricardo} where discrete
solitons were constructed numerically using a dynamical reduction. 
Dark solitons and staggered solutions of the
cubic--quintic DNLS equation were recently studied numerically in Refs.~\cite{Belgrade1}
and \cite{Belgrade2} respectively.

We consider a discrete nonlinear Schr\"odinger equation with a
cubic--quintic nonlinearity in the form,
\begin{equation}
i  \dot{ \psi }_n + C ( \psi_{n+1} + \psi_{n-1} - 2 \psi_n ) + B |
\psi_n |^2 \psi_n - Q | \psi_n |^4 \psi_n =0, \quad n \in
\mathbb{Z}, \label{cqdnls}
\end{equation}
where $\psi_n(t) : \mathbb{R}_+ \to \mathbb{C}$ and $(C,B,Q)$ are
real-valued parameters. Nonlinear Schr\"odinger lattices have
proved to be relevant models in a variety of contexts (see reviews
in \cite{EJ03,KRB01}), including the description of optical pulses
in one-dimensional waveguide arrays \cite{Demetri}. In this
application, the quantity $|\psi_n|^2$ represents the intensity of
the electric field of waveguide $n$, $C>0$ represents coupling
strength between adjacent waveguides and $(B,Q)$ measure the
nonlinearity strength. A large portion of the literature is
dedicated to Eq.~\eqref{cqdnls} with $Q=0$, which would correspond
to a medium with a Kerr nonlinearity. Recent experimental results
\cite{CQ1,CQ3,CQ2} have shown that the nonlinear response of some materials is better
fit with an additional competing quintic nonlinearity, i.e. $B,Q >
0$. This lends relevance to studying the
cubic-quintic DNLS equation in the form (\ref{cqdnls}).

The cubic--quintic DNLS equation (\ref{cqdnls}) has two conserved
quantities, namely the power,
\begin{equation}
M = \sum_{n \in \mathbb{Z}} |\psi_n|^2, \label{norm}
\end{equation}
and the Hamiltonian,
\begin{equation}
H = \sum_{n\in \mathbb{Z}}   C \left( \psi^*_n \psi_{n+1} +
\psi_n \psi^*_{n+1} - 2 |\psi_n|^2 \right)  + \frac{B}{2}|
\psi_n| ^4 - \frac{Q}{3}|\psi_n| ^6. \label{ham}
\end{equation}
Steady-state solutions have the form $\psi_n = u_n e^{-i \mu t}$,
$n \in \mathbb{Z}$, where $\mu \in \R$ and $u_n \in \mathbb{R}$
are found from the stationary DNLS equation,
\begin{equation}
 \mu u_n + C ( u_{n+1} + u_{n-1} - 2 u_n ) +  B  u_n^3 - Q u_n ^5
=0, \quad n \in \mathbb{Z}. \label{cqdnls-sta}
\end{equation}
We seek localized solutions of the stationary DNLS equation
\eqref{cqdnls-sta} which in turn correspond to \emph{discrete
solitons}.

Spectral stability of the steady-state solutions is studied with
the linearization ansatz,
$$
\psi_n(t) = \left( u_n + (v_n + i w_n) e^{\lambda t} + ( v^*_n
+ i w^*_n ) e^{\lambda^* t} \right) e^{-i \mu t}, \quad n
\in \mathbb{Z},
$$
which leads to the spectral problem,
\begin{equation}
\label{eigenvalue-problem} \left\{ \begin{array}{cc} -\mu v_n -C(
v_{n+1} + v_{n-1} - 2v_n) -
3 B u_n^2 v_n + 5 Q u_n^4 v_n = -\lambda w_n, \\
-\mu w_n -C( w_{n+1} + w_{n-1} - 2w_n) - B u_n^2 w_n + Q u_n^4 w_n
= \lambda v_n, \end{array} \right.  \quad n \in \mathbb{Z}.
\end{equation}
We are looking for nonzero solutions of the linearized system in
$L^2(\mathbb{Z},\mathbb{C}^2)$ called eigenvectors. The
steady-state solution is called
unstable if there exists at least one eigenvector for which ${\rm
Re}(\lambda) > 0$. Otherwise, the solution is called spectrally
stable.

Steady-state solutions are considered in Section 2. Spectral stability of
these solutions is studied in Section 3. For both problems,
we compare results of the variational approximations and the direct
numerical approximations.

\section{Variational approximations of the steady-state solutions}

Solutions of the DNLS equation \eqref{cqdnls} correspond to
critical points of the Lagrangian,
\begin{equation}
\L = \sum_{n \in \mathbb{Z}} \frac{i}{2} \left( \psi^*_n \pa_t
\psi_n - \psi_n \pa_t \psi^*_n \right) + C \left( \psi^*_n
\psi_{n+1} + \psi_n \psi^*_{n+1} - 2 |\psi_n|^2 \right) +
\frac{B}{2}| \psi_n| ^4 - \frac{Q}{3}|\psi_n| ^6. \label{lag}
\end{equation}
To find approximate solutions for discrete solitons, we pose a
trial function in the form,
\begin{equation}
 \psi_n = A e^{ i \phi_n } e^{ -\eta |n - n_0 |},
 \quad \phi_n = \alpha + k(n-n_0) + \frac{\beta}{2}(n - n_0)^2,
\label{ansatz}
\end{equation}
where each of the six parameters are dependent on $t$.
Substituting Eq.~\eqref{ansatz} into the Lagrangian \eqref{lag}
and evaluating the sums yields the effective Lagrangian,
\begin{eqnarray}
\L_{\mathrm{eff}} & = & -A^2 \left(\da - \frac{k}{2}\dc \right)S_0
-A^2 \left(\dk - \frac{\beta}{2}\dc \right)S_1 -\frac{A^2}{2}\db
S_2 \nonumber \\ & \phantom{t} &  + C A^2 \left( e^{ik}S_\beta +
e^{-ik}S_\beta^* - 2 S_0 \right) + \frac{B}{2}A^4 S_4 -
\frac{Q}{3}A^6S_6, \label{leff}
\end{eqnarray}
where $\chi=2n_0-1$ and,
\begin{eqnarray*}
 S_0(\eta,\chi) &=& {\frac {\cosh \chi\eta  }{\sinh \eta  }}, \\
 S_1(\eta,\chi) &=& \frac{ \cosh\eta \sinh \chi\eta  }{2 \sinh^2 \eta}  + \frac{\chi}{2} S_0,  \\
 S_2(\eta,\chi)  &=& \left (\frac{2}{\sinh^2 \eta }  + \frac{1}{4} \right ) S_0 - \frac{\chi}{2} \frac{ \cosh\eta \sinh \chi\eta  }{\sinh^2 \eta} - \frac{\chi^2}{4} S_0, \\
 S_4(\eta,\chi)  &=&  S_0(2\eta,\chi), \\
 S_6(\eta,\chi)  &=&  S_0(3\eta,\chi), \\
S_{\beta}(\eta,\chi,\beta)&=& e^{-\eta}e^{- \frac{i}{2} \beta \chi}
                \left( 1 + \frac{e^{-\eta \chi}}{e^{i \beta}e^{\eta}-e^{-\eta} }
        +                  \frac{e^{\eta \chi}}{e^{-i \beta}e^{\eta}-e^{-\eta}}
        \right).
\end{eqnarray*}
Since the center $n_0$ of the ansatz (\ref{ansatz}) can be
arbitrarily chosen on $[0,1]$ module the discrete group of
translations in $n \in \mathbb{Z}$, we shall consider $\chi$
on $[-1,1]$ (this restriction was used already in the derivation
of (\ref{leff})). The solution with $\chi=0$ is centered between
lattice sites and hence is called \textit{bond-centered}. The
solution with $\chi=\pm 1$ is centered on a lattice site and hence
is called \textit{site-centered}. Solutions for $\chi \in (-1,0)
\cup (0,1)$ are called \textit{asymmetric}.

According to the variational principle, the
effective Lagrangian $\L_{\mathrm{eff}}$ achieves critical values
at the Euler-Lagrange equations,
\begin{equation}
\label{Euler-Lagrange} \frac{\pa \L_{\mathrm{eff}} }{\pa p_j} -
\frac{d}{dt} \left[ \frac{\pa \L_{\mathrm{eff}}}{\pa \dot{p}_j}
\right]=0,
\end{equation}
where $p_j$ represents a parameter of ansatz \eqref{ansatz}.
Varying $\alpha$ yields the conservation law,
\begin{equation}
 A^2 S_0 = M,
\label{va1}
\end{equation}
which corresponds to the dynamical invariant \eqref{norm} of the
power. Varying $A$ yields,
\begin{eqnarray}
 \da &=& \dc \frac{k}{2}  -  \dk \frac{S_1}{S_0}  + \frac{\beta}{2}\frac{S_1}{S_0}\dc -
 \frac{1}{2}\db \frac{S_2}{S_0}   + \frac{C}{S_0}( e^{ik}S_{\beta}+e^{-ik}S_{\beta}^*-2S_0) \nonumber \\
& \phantom{t} & \; + B M \frac{S_4}{S_0^2} - Q M^2 \frac{S_6}{S_0^3} \; .
\label{va2}
\end{eqnarray}
\begin{figure}[t]
\centerline{
 \epsfig{file=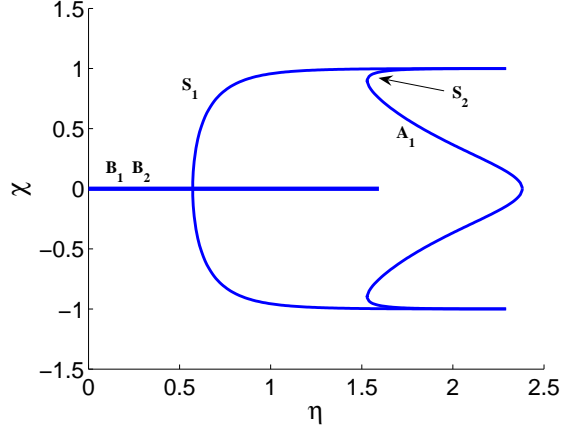,width=8cm,angle=0}}
\caption{Solutions of Eqs.~\eqref{dchi} and  ~\eqref{deta} for
$C=0.1$, $B = 2$, and $Q = 1$. } \label{chi_eta_bif}
\end{figure}
Before writing the remaining equations, it will be more convenient
to make use of the fact that we seek steady-state solutions. This
corresponds to
$$
\beta=k=\dc=\de=0 \quad \mbox{\rm and} \quad \frac{d\alpha}{d t} =
-\mu,
$$
where $\mu$ is the parameter of the steady-state solution. With this assumption,
the equations corresponding to variation of $k$ and $\beta$ are
identically satisfied. Varying $\chi$ and $\eta$ and making use of
Eq.~\eqref{va2} leads to the following two equations respectively,
\begin{equation}
\frac{ A^2 \eta \sinh \eta \chi }{ \sinh \eta \cosh \eta \chi   }
\left(  \frac{A^4 2Q\cosh \eta \cosh \eta \chi}{4 \cosh^2 \eta -1 }    -A^2
 + C e^{-\eta}\sinh 2\eta    \right) =0,
 \label{dchi}
\end{equation}
and,
\begin{eqnarray} \label{deta}
& \phantom{t} & -C e^{-\eta } \left(1+ \frac{\cosh\eta   \chi - \cosh\eta }{\sinh\eta } +
 \frac{ \chi \sinh\eta   \chi}{\cosh\eta   \chi} \right)
 +   \frac{B A^2 \chi \sinh\eta \chi }{2 \sinh2\eta  \cosh\eta   \chi } \\ \nonumber  
 & \phantom{t} & + \frac{B A^2 \cosh2 \eta\chi}{2 \sinh^2 2 \eta }
 -  Q  A^4 \left(  \frac{  \chi \sinh\eta   \chi}{\sinh3 \eta  }
 + \frac{  \cosh\eta  \cosh3 \eta   \chi}{\sinh^23 \eta } \right)=0.
\end{eqnarray}
Note these equations  with $Q=0$ correspond to those in
Ref.~\cite{Ka05}. When $\chi=0$ (bond-centered solutions),
Eq.~\eqref{dchi} is identically satisfied and Eq.~\eqref{deta}
becomes,
\begin{eqnarray}
  \frac{ - C e^{-\eta } (1+ e^{-\eta}) \sinh^2 2\eta   }{\sinh \eta}
 + \frac{B A^2}{2}  -  \frac{Q A^4  \cosh\eta \sinh^2 2 \eta}{\sinh^2 3 \eta}
=0,
\label{chi0}
\end{eqnarray}
which is easily solved for $A^2$, giving an existence condition
in terms of the parameter $\eta$. There exists exactly two
bond-centered solutions (denoted by $B_1$ and $B_2$) for $\eta \in (0,\eta_{\rm cr})$, which
disappear as a result of the saddle--node bifurcation at $\eta =
\eta_{\rm cr}$. See Fig.~\ref{chi_eta_bif}, where solutions
of Eqs.~\eqref{dchi} and \eqref{deta} are plotted for $C=0.1$, $B = 2$,
and $Q = 1$ with $\eta_{\rm cr}\approx 1.56$. We note, for larger $C$
steady-state solutions become smoother (they approach the continuous
counterpart) and so the ansatz (\ref{ansatz}), which is based on an exponential cusp,
becomes irrelevant.

For $\chi \neq 0$, the term in parenthesis of  Eq.~\eqref{dchi}
can be used as condition for $A^2$, which can then be substituted
into \eqref{deta} yielding a root finding problem in $(\eta,\chi)$
parameter space. We find exactly three pairs of solutions for $C = 0.1$, $B = 2$,
and $Q = 1$, which are denoted by $S_1$,$S_2$, and $A_1$ in
Fig.~\ref{chi_eta_bif}. Branches $(S_1,S_2)$ and $(S_2,A_1)$ are connected
by means of the saddle-node bifurcations. Branch $S_1$ arises from
$\chi = 0$ as a result of the supercritical pitchfork bifurcation, while branches
$S_2$ and $A_1$ arise from $\chi = 0$ as a result of the subcritical pitchfork
bifurcation.

The solution branches $S_1$ and $S_2$ approach $\chi =1$ rapidly, so that they correspond to
the site-centered solitons slightly disturbed by the variational
approximation. The solution branch $A_1$ is a true asymmetric
solution that connects the bond-centered and site-centered
solitons. Only one branch $S_1$ of site-centered solutions and only one branch $B_1$ of
bond-centered solutions exist in the cubic case $Q = 0$, where
these two branches extend for any $\eta > 0$.

\begin{figure}[tbh]
\centerline{
 \epsfig{file=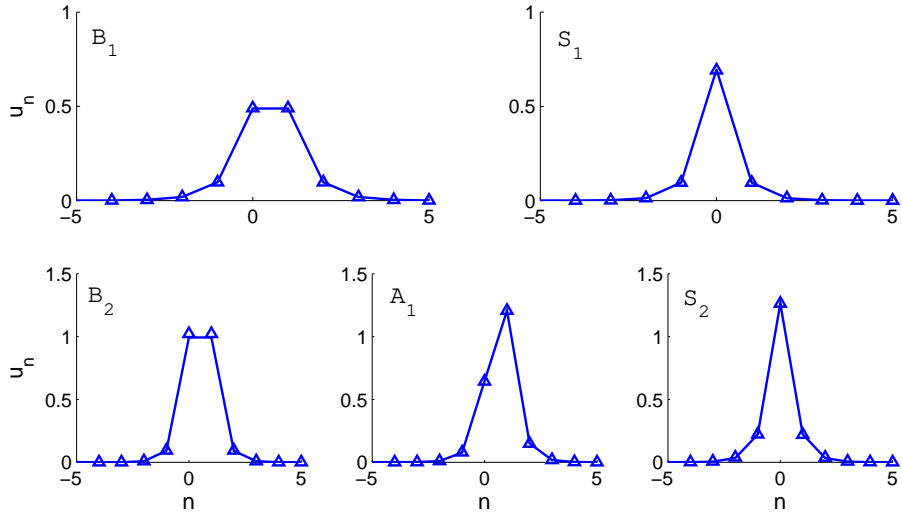,width=14cm,angle=0}}
\caption{Plot of numerical (solid lines) and variational
(markers) solutions for $C=0.1$, $B=2$, $Q=1$. Top: Short
bond-centered (left) and site-centered (right) solutions. Bottom:
A taller bond-centered (left) and site-centered (right) solution.
The asymmetric solution (middle) is an intermediate between
the two symmetric profiles. The labels
$A_1$, $B_1$, $B_2$, $S_1$, and $S_2$ correspond to those in Fig.~\ref{chi_eta_bif}. }
\label{profiles}
\end{figure}

To test the validity of the variational approximations we compare
them against direct numerical solutions of the stationary DNLS equation
(\ref{cqdnls-sta}). The ``numerically exact" solutions are obtained using a Newton
method with the VA solutions as initial seeds.
The profiles of the discrete solitons that are obtained via the VA and the corresponding
numerical solutions are shown in Fig.~\ref{profiles}.

\begin{figure}[thb]
\centerline{
 \epsfig{file=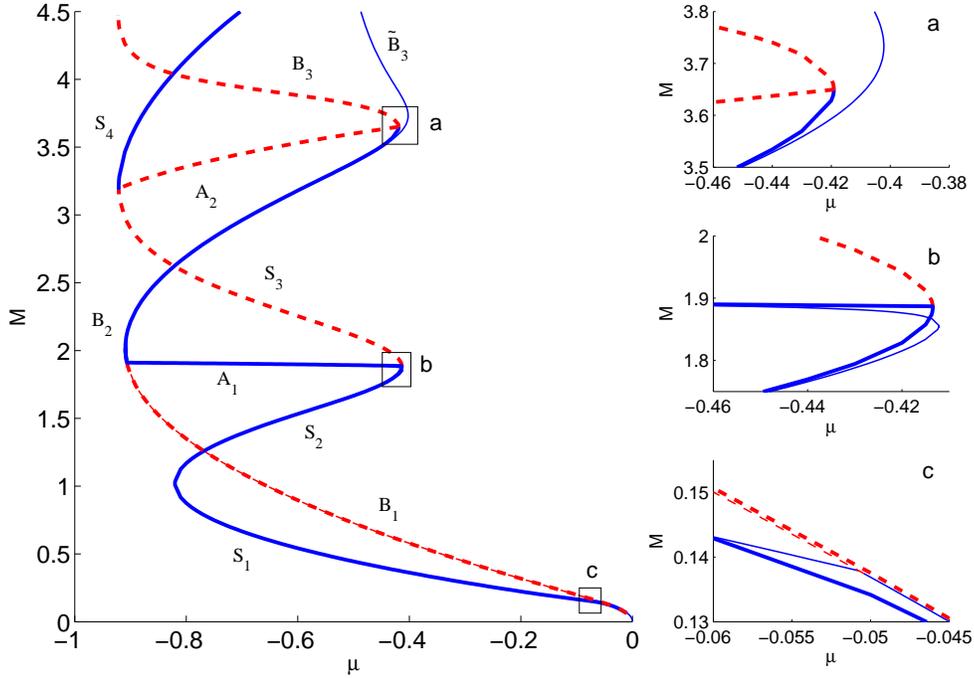,width=15cm,angle=0}}
\caption{(Color online)  Plot of the power $M$ of the VA solutions (thin lines) and numerical solutions
(thick lines). Stable branches are represented by solid blue lines
and unstable branches by dashed red lines.
There are two principle branches, one corresponding to site-centered
solutions (labeled $S_j, \, j=1..4$) and one for bond-centered solutions
(labeled $B_j, \, j=1..3$). The two principle branches are connected
via asymmetric solutions (labeled $A_j,\, j=1,2$) at points of
stability change. Only the branches $S_1,S_2,A_1,B_1,$ and $B_2$,
are captured by the VA. Differences between the VA
and numerical solutions are more visible
in the zooms labeled `a', `b' and `c'. The branch labeled
$\tilde{B}_3$ corresponds to the VA that fails to capture $B_3$. } \label{big}
\end{figure}
\begin{figure}[tbh]
\centerline{
 \epsfig{file=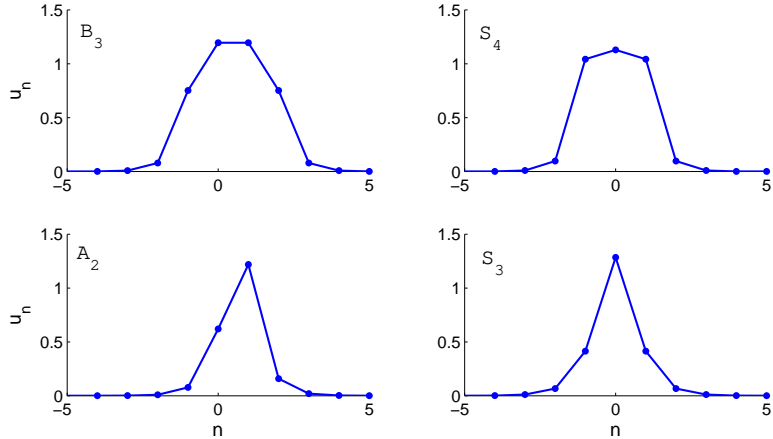,width=12cm,angle=0}}
\caption{Examples of numerical solutions
shown in Fig.~\ref{big} that the VA did not capture. The failure
of the VA is due to the wide nature of the solutions. Parameter
values are $C=0.1$, $B=2$, and $Q=1$.}
\label{profiles_nova}
\end{figure}
It is more instructive for comparison to plot the solution branches in the $(\mu,M)$
plane, where $M$ is defined in Eq.~\eqref{va1} and $\mu = -\frac{d \alpha}{d t}$,
see Fig.~\ref{big}. The predicted stability is also depicted (see Sec.~\ref{stability} for details
on stability computations).

Besides the five solutions predicted within the variational
approximation, there exists additional branches $S_3$, $S_4$, $B_3$, $A_2$ of
solutions of  Eq. (\ref{cqdnls-sta}) that appear on Fig.~\ref{big}.
See Fig.~\ref{profiles_nova} for profiles of these ``wide" solutions.
The pattern of Fig.~\ref{big} suggests existence of an infinite number of additional branches of discrete
solitons for large power $M$ with a characteristic snaking
behavior. The snaking behavior was also found in the higher dimensional
cubic-quintic DNLS equation \cite{Ch09} and the Swift-Hohenberg equation \cite{Loyd2,Loyd1}.
It seems the role a higher order dispersion plays
in the Swift-Hohenberg equation is replaced by discreteness in the
cubic-quintic DNLS equation.

\begin{figure}[tbh]
\centerline{
 \epsfig{file=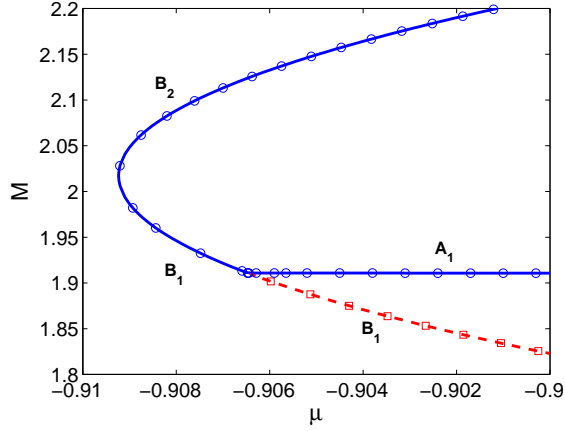,width=8cm,angle=0}}
\caption{Plot of numerical (lines) and variational
(markers) approximations of power $M$ versus $\mu$ 
for $C=0.1$, $B=2$, $Q=1$ in a parameter region
 near a pitchfork bifurcation. Stable solutions of
the VA are indicated by circles and unstable solutions by squares.
} \label{pitchfork}
\end{figure}

The bifurcations shown in Fig.~\ref{big} look as if the asymmetric
solutions connect exactly at the turning points of bond-centered
or site-centered solutions. This is
not the case. Rather, there exists a pair of
saddle--node and pitchfork bifurcations and the stability exchange occurs
at the pitchfork bifurcation where the asymmetric solutions emanate.
Interestingly, the VA is able to accurately capture this subtle
bifurcation scenario. In Fig.~\ref{pitchfork} a plot of such
a bifurcation pair is shown. This particular pair represents
a saddle-node bifurcation of a short ($B_1$) and tall ($B_2$)
bond-centered solution and a pitchfork of two asymmetric solutions
($A_1$) and the short bond-centered solution ($B_1$). The agreement between the
variational and numerical approximations is impressive.
The stability is also correctly predicted.

On the other hand, bifurcations including the site-centered
solutions are not captured as well. This should not be surprising
since site-centered solutions are never \textit{truly} represented
by the VA
(see Fig.~\ref{chi_eta_bif}). We point out other minor discrepancies between the variational
approximations and the numerical results in the zooms of Fig.~\ref{big}.
Zoom (a) shows where the VA solution $\tilde{B}_3$ departs from the corresponding
numerical solution $B_3$. This is expected
as the ansatz \eqref{ansatz} is only valid for ``narrow" solutions,
i.e. those that have at most two initially excited sites.
Zoom (b) shows that the asymmetric solution $A_1$
is connected to the site-centered solution $S_2$ (opposed to $S_3$) and is underestimated.
The wider short site-centered solution $S_3$ is not captured at all by the VA (as expected).
Finally, zoom (c) shows that the VA falsely predicts collision of the bond-centered $B_1$ and
site-centered $S_1$ solutions for a non-zero value of $M$, similar to the
cubic case discussed by Kaup in \cite{Ka05}.

\section{Variational approximations of spectral stability}
\label{stability}

In order to determine stability within the VA we return to the
variational equations (\ref{Euler-Lagrange}) but this time without
the assumption that $\beta=k=\dc=\de=0$. Let
$\vec{x}=(\beta,\chi,\eta,k)^T$ represent the four parameters of
ansatz (\ref{ansatz}) after Eqs.~(\ref{va1}) and (\ref{va2}) are
used. To perform a linear stability analysis we substitute,
\begin{equation*}
\vec{x} = \vec{x}_0 + \epsilon \vec{y} e^{\lambda t},   \quad \lambda \in \mathbb{C},
\end{equation*}
into the four variational equations, where the steady-state solution
is defined by $\vec{x}_0= (0,\chi_0,\eta_0,0)^T$ and
$(\chi_0,\eta_0)$ satisfy Eqs.~\eqref{dchi} and \eqref{deta}.
Keeping only the terms linear in $\epsilon$ leads to the
generalized eigenvalue problem,
\begin{equation}
\label{generalized-eigenvalue}
\lambda {\bf A} \vec{y} = {\bf B} \vec{y},
\end{equation}
where the entries of the $4 \times 4$ matrices  ${\bf A}$ and ${\bf B}$ are
given by,
\begin{align*}
a_{11}&= 0,&
a_{21}&=  0,& \\
a_{12}&= \frac{M}{2} + M \pc \left[ \frac{S_1}{S_0} \right],&
a_{22}&=\frac{M}{2}\left( \frac{S_1}{S_0}  +  \pc \left[ \frac{S_2}{S_0} \right] \right), &  \\
a_{13}&= M \pe \left[ \frac{S_1}{S_0} \right],&
a_{23}&=  \frac{M}{2} \pe \left[ \frac{S_2}{S_0} \right],& \\
a_{14}&=  0,&
a_{24}&=  0, &  \\
\phantom{a_{14}} &\phantom{=}  \phantom{0}& \phantom{a_{24}}&\phantom{=}  \phantom{0} &  \\
a_{31}&= \frac{M}{2 S_0}\left( \frac{\pa S_0}{\pa \chi} \frac{S_2}{S_0}
            - \frac{\pa S_2}{\pa \chi}  - S_1 \right),   &
a_{41}&=   \frac{M}{2 S_0}\left( \frac{\pa S_0}{\pa \eta} \frac{S_2}{S_0}
                  - \frac{\pa S_2}{\pa \eta} \right),  & \\
a_{32}&= 0,&
a_{42}&=0,&  \\
a_{33}&= 0,&
a_{43}&=  0,& \\
a_{34}&=   \frac{M}{ S_0}\left( \frac{\pa S_0}{\pa \chi} \frac{S_1}{S_0}
              - \frac{\pa S_1}{\pa \chi} \right) - \frac{M}{2} ,&
a_{44}&=   \frac{M}{ S_0}\left( \frac{\pa S_0}{\pa \eta} \frac{S_1}{S_0}
                - \frac{\pa S_1}{\pa \eta} \right), &
\end{align*}
and,
\begin{align*}
b_{11}&= \frac{C M}{S_0}\left( \frac{\sinh \eta\chi }{\sinh^2 \eta}
- \chi e^{- \eta}(1 + S_0)\right),&
b_{21}&= \frac{-C M}{S_0}\left( \frac{\pa^2 S_\beta}{\pa \beta^2} + \frac{\pa^2 S_\beta^*}{\pa \beta^2}    \right)  ,& \\
b_{12}&= 0,&
b_{22}&=  0,& \\
b_{13}&=0,&
b_{23}&=  0,& \\
b_{14}&=  \frac{2 C M}{S_0}e^{- \eta}(1 + S_0),&
b_{24}&=  \frac{C M}{S_0}\left( \frac{\sinh \eta\chi }{\sinh^2 \eta}
- \chi e^{- \eta}(1 + S_0)\right), &  \\
\phantom{a_{14}} &\phantom{=}  \phantom{0}& \phantom{a_{24}}&\phantom{=}  \phantom{0} &  \\
b_{31}&= 0, &
b_{41}&= 0, & \\
b_{32}&=  - \frac{\pa }{\pa \chi} \left[ \frac{\pa L}{\pa \chi} + \frac{\pa S_0}{\pa \chi} P \right],&
b_{42}&= -  \frac{\pa }{\pa \chi} \left[ \frac{\pa L}{\pa \eta} + \frac{\pa S_0}{\pa \eta} P \right],&  \\
b_{33}&= -  \frac{\pa }{\pa \eta} \left[ \frac{\pa L}{\pa \chi} + \frac{\pa S_0}{\pa \chi} P \right],&
b_{43}&= -  \frac{\pa }{\pa \eta} \left[ \frac{\pa L}{\pa \eta} + \frac{\pa S_0}{\pa \eta} P \right],& \\
b_{34}&= 0 ,&
b_{44}&= 0.&
\end{align*}
Here we have defined,
\begin{eqnarray*}
L&=& \frac{C M}{S_0}\left( S_\beta + S^*_\beta -2 S_0\right) + \frac{B M^2}{2 S_0^2} S_4
- \frac{Q M^3}{3 S_0^3} S_6, \\
P&=& \frac{-C M}{S_0^2}\left( S_\beta + S^*_\beta  - 2 S_0\right) - \frac{B M^2}{ S_0^3} S_4
+ \frac{Q M^3}{ S_0^4} S_6,
\end{eqnarray*}
and each entry is evaluated at the fixed point $\vec{x}_0$. For
the form of the perturbation chosen, an eigenvalue with Re$(\lambda) > 0$
indicates instability of the steady-state solution $\vec{x}_0$.

\begin{figure}[thb]
\centerline{
 \epsfig{file=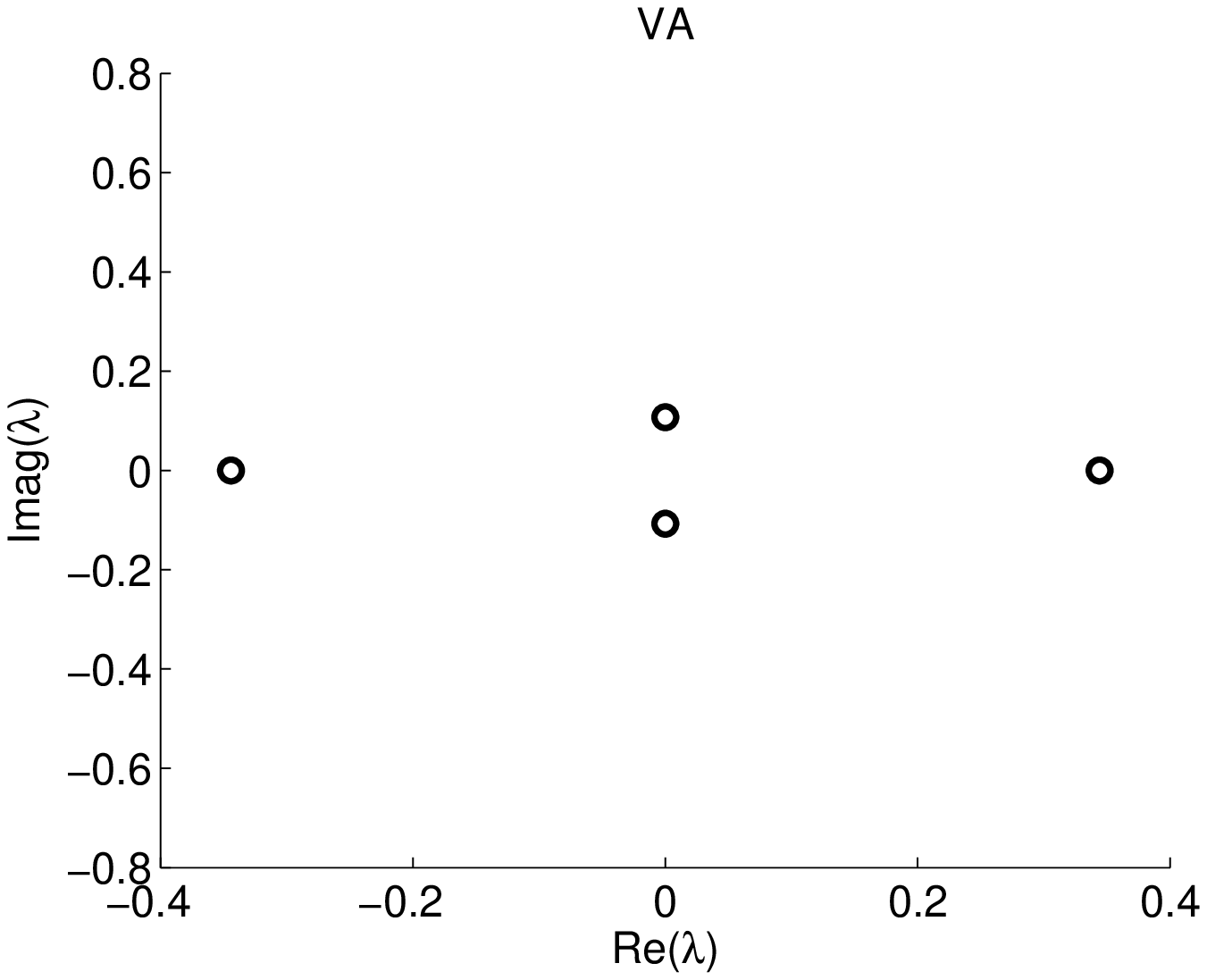,width=7cm,angle=0}
  \epsfig{file=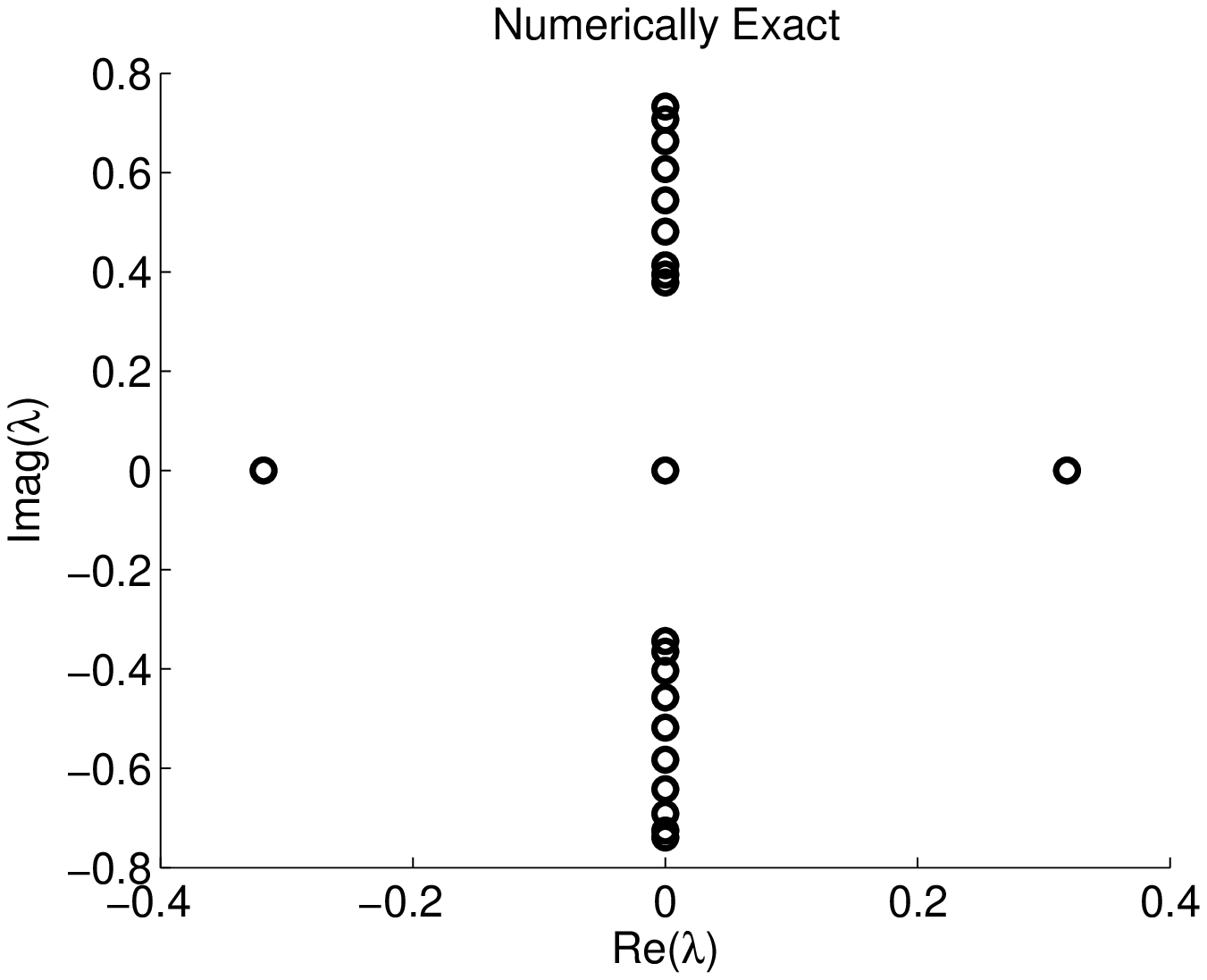,width=7cm,angle=0} }
\caption{Eigenvalues corresponding to the bond-centered solution $B_1$
in the variational system (left) and the full system (right). } \label{quad}
\end{figure}

Since ${\bf A}$ and ${\bf B}$ are real-valued matrices, both $\lambda$ and $\lambda^*$ 
are eigenvalues. Since ${\bf A}$ and ${\bf B}$ has a clear block structure, if
$\lambda$ is an eigenvalue with the eigenvector $\vec{y}$, then
$-\lambda$ is an eigenvalue with the eigenvector $S \vec{y}$, where
$S = {\rm diag}(1,-1,-1,1)$. Therefore, eigenvalues of (\ref{generalized-eigenvalue})
occur as pairs of real or imaginary eigenvalues or as quartets of complex eigenvalues.
A typical example of eigenvalues of (\ref{generalized-eigenvalue}) is shown on the left
panel of Fig. \ref{quad} for the unstable solution $B_1$ with a pair of real and a pair of
imaginary eigenvalues.

\begin{figure}[thb]
\centerline{
 \epsfig{file=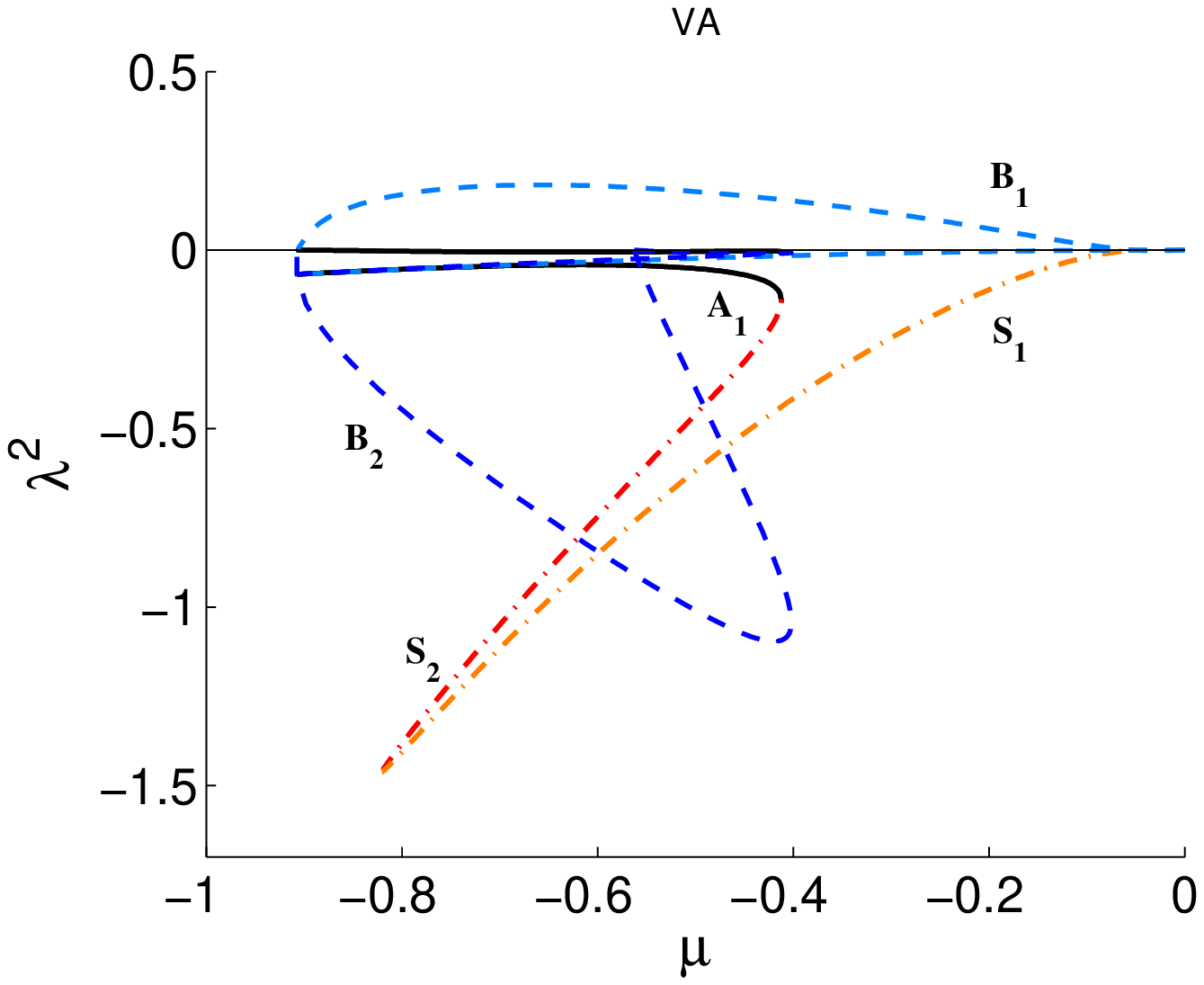,width=7cm,angle=0}
  \epsfig{file=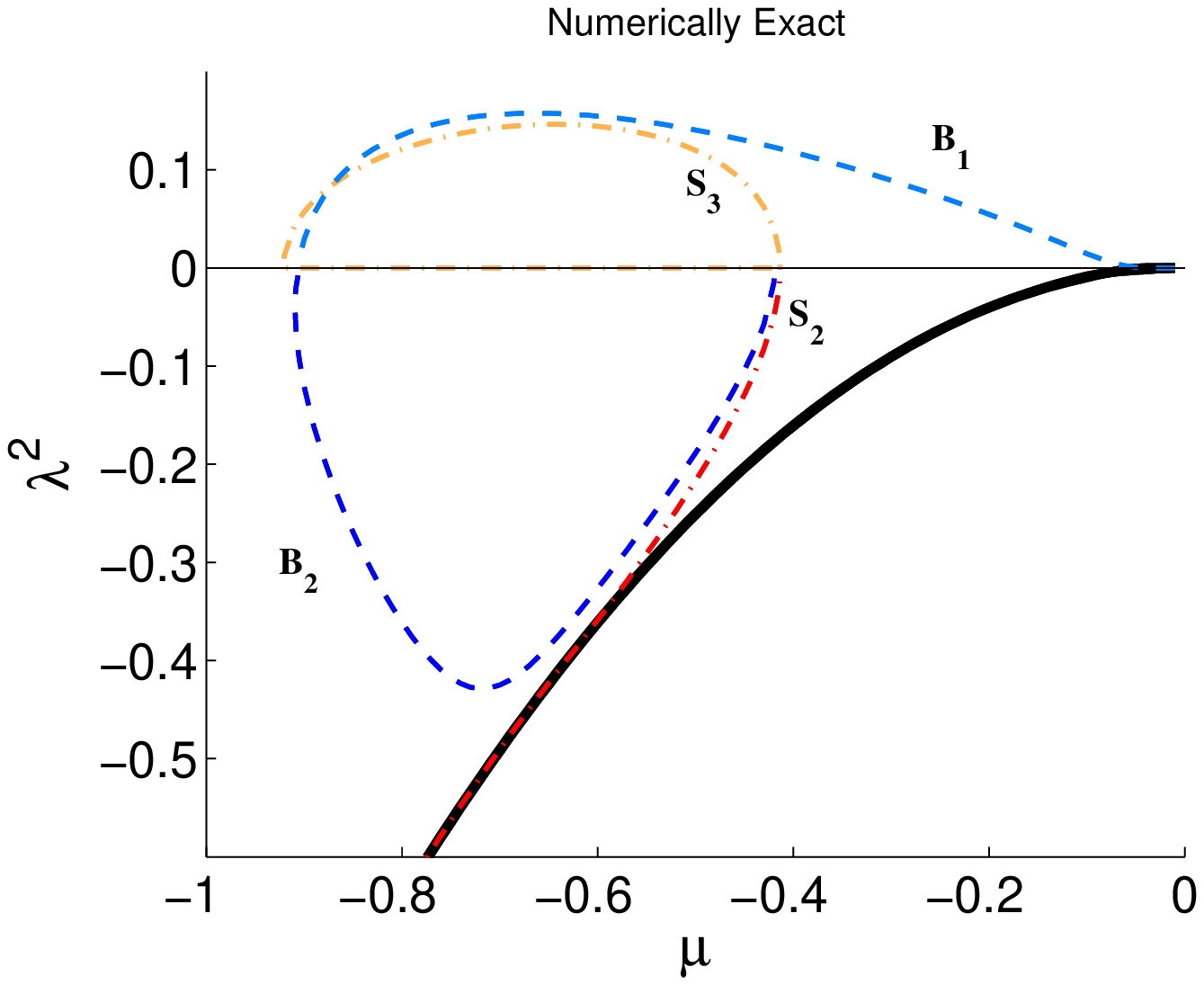,width=7cm,angle=0} }
 \centerline{
 \epsfig{file=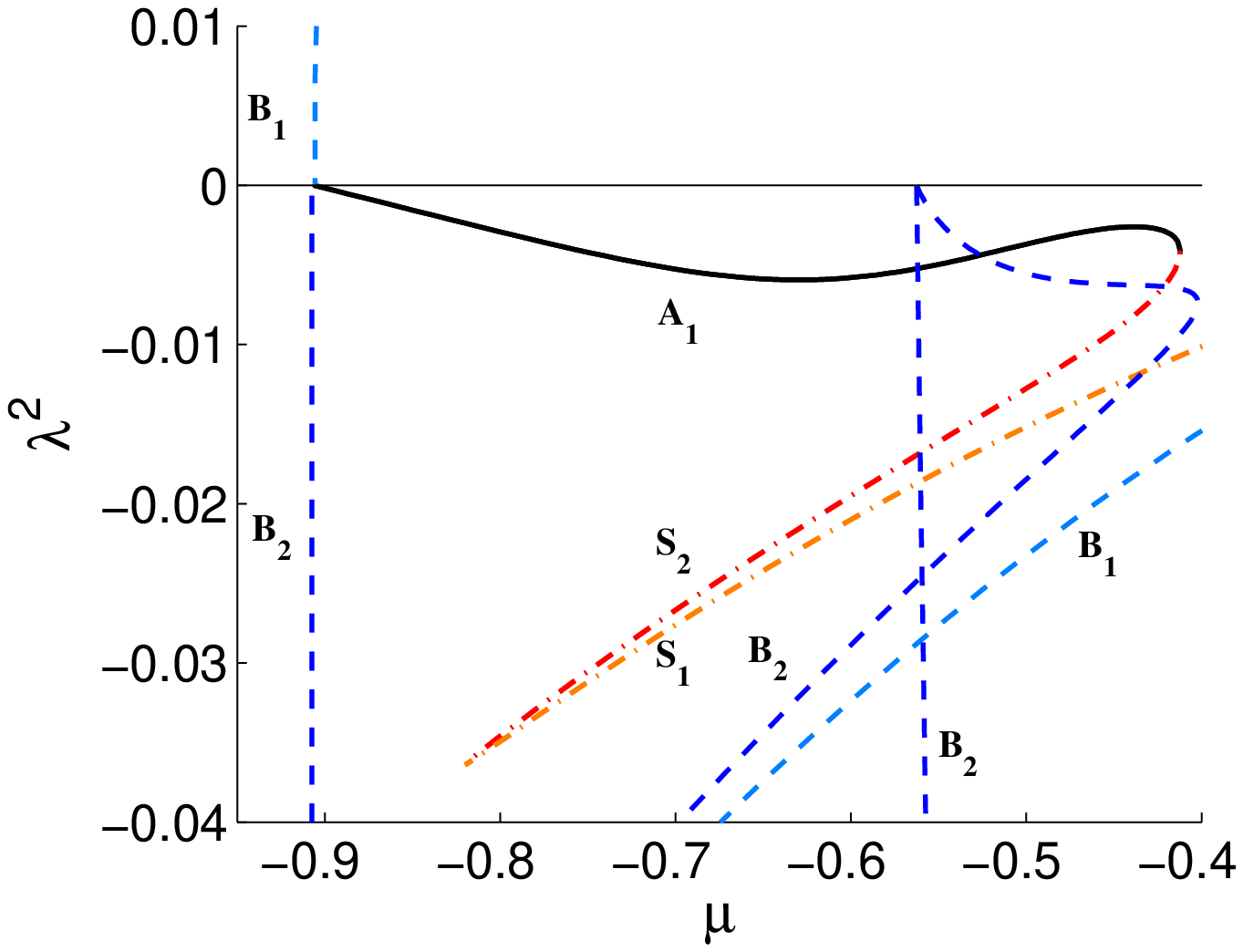,width=7cm,angle=0}
 \epsfig{file=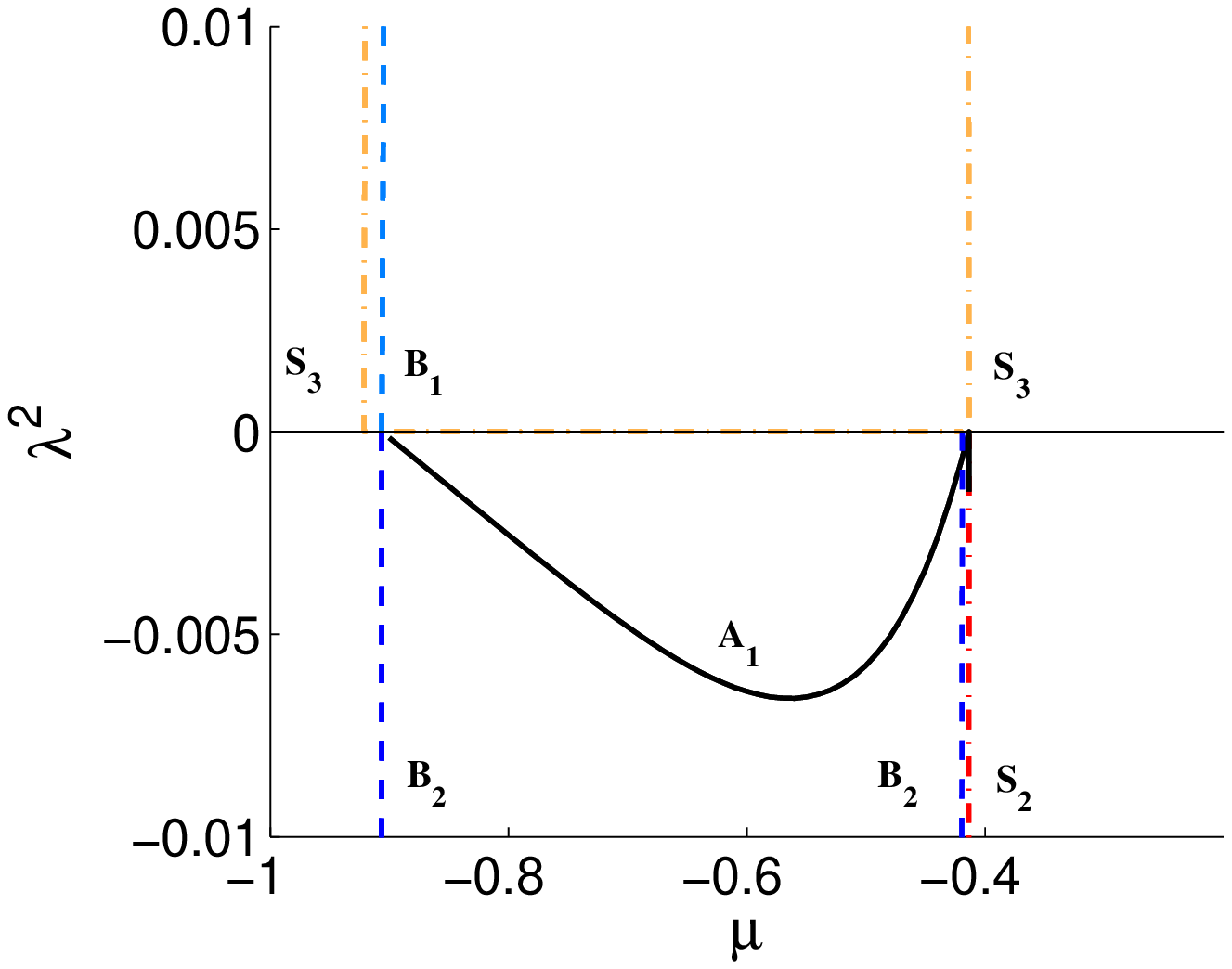,width=7cm,angle=0} }
\caption{(Color online) Top: plot of eigenvalues for the VA solutions
(left) and numerical solutions (right) at $C=0.1$, $B = 2$, and 
$Q = 1$. Site-centered branches are plotted as
dashed-dot lines (in various shades of red for each branch) and
bond-centered branches as dashed lines (in various shades of blue).
The asymmetric solution is shown as a solid black line. 
The thick solid black line shows the boundary of the continuous spectral band 
in the full problem. Bottom: zooms of the top panels
near $\lambda^2 \approx 0$. }
\label{spectrum}
\end{figure}

The stability of the VA solutions corresponding to Fig.~\ref{big}
was computed from eigenvalues of the generalized problem (\ref{generalized-eigenvalue}).
The spectrum for each of the variational
solutions shown in Fig.~\ref{big} is plotted in the left panels of
Fig.~\ref{spectrum}. Only the bond-centered solution $B_1$ has a branch
that is predicted to be unstable. The bottom left panel of Fig.~\ref{spectrum}
is a zoom of small eigenvalues in the
top left panel. Where the asymmetric solution $A_1$ meets the
bond-centered solutions $B_1$ and $B_2$ corresponds to stability exchange, whereas
the connection to the site-centered solution $S_2$ occurs at $\lambda^2
< 0$ and hence no stability change takes place. Note that each branch of VA solutions
has exactly two pairs of real or imaginary eigenvalues of the generalized
problem (\ref{generalized-eigenvalue}).

A linear stability analysis of the ``numerically exact'' solutions was
also carried out from the spectral stability problem (\ref{eigenvalue-problem})
using the procedure described in Ref.~\cite{Ricardo}. The numerical approximation
of the spectrum for the unstable bond-centered solution $B_1$ is
shown on the right panel of Fig. \ref{quad}. Note that there is always a double zero
eigenvalue in the stability problem (\ref{eigenvalue-problem}) which is related to
the gauge symmetry. This double zero eigenvalue corresponds to the eigenvector
$(v_n,w_n) = (0,u_n)$ and the generalized eigenvector
$(v_n,w_n) = (\partial_{\mu} u_n, 0)$ of system (\ref{eigenvalue-problem}).
This double zero eigenvalue is captured by the variational approximation and
it results in the conservation law (\ref{va1}) and the variational equation (\ref{va2}).
On the other hand, the VA gives two pairs of eigenvalues and only the real pair 
persists in the numerical approximation. The purely imaginary pair of eigenvalues does not exist 
in the numerical solutions, where it is replaced by the continuous spectral band on 
the two symmetric intervals of the purely imaginary axis.

Isolated, non-zero eigenvalues for solutions $B_1$, $B_2$, $S_2$, $S_3$, and $A_1$
are shown in the right panels of Fig.~\ref{spectrum}. We note that
no isolated eigenvalues of $S_1$ exist. The overall ``look" of the top
panels of Fig.~\ref{spectrum} are quite similar, but more
importantly, the stability is correctly predicted. Note that the right panels of 
Fig. \ref{spectrum} contains fewer eigenvalues
than the left panels of Fig. \ref{spectrum} because some of the purely imaginary eigenvalues
of the VA solutions approximate the continuous spectrum of the numerical solutions.
An interesting difference is seen in the bottom right panel of Fig. \ref{spectrum}. In the
full problem, the eigenvalue pair for the asymmetric solution $A_1$ vanishes 
when $A_1$ intersects with the site-centered solution $S_2$. 

In conclusion, we have extended the results of Ref.~\cite{Ka05} to
show that not only does the VA faithfully represent the
fundamental localized modes, but is also able to correctly predict the
corresponding stability for small coupling constant $C$ and power $M$.
We showed this in the context of the cubic--quintic DNLS equation,
which exhibits a family of discrete solitons,
five of which were accurately captured by the variational
approximation. It would be interesting to derive the variational
equations in the context of time-dependent perturbations, although,
the resulting equations would be far more complex and may
undermine the utility of the method. It would also be relevant to
extend the results of \cite{Ch09} and apply this analysis to the
higher dimensional cubic--quintic DNLS equation.

\section*{Acknowledgements} C.C. would like to thank R.
Carretero-Gonz\'alez, D.J. Kaup, P.G. Kevrekidis, and B.A.
Malomed, for insightful discussions. C.C. also
appreciates the hospitality at the Department
of Mathematics at McMaster University.

\medskip
Received xxxx 20xx; revised xxxx 20xx.
\medskip


\begin{thebibliography}{99}
\bibitem{Loyd2} M. Beck, J. Knobloch, D.J.B. Lloyd, B. Sandstede, T. Wagenknecht,
\emph{Snakes, ladders, and isolas of localised patterns}, SIAM J. Math. Anal, accepted.

\bibitem{CQ1} G. Boudebs, S.Cherukulappurath, H. Leblond, J. Troles, F. Smektala, F. Sanchez,\emph{ Experimental and theoretical study of higher-order nonlinearities in chalcogenide glasses}, Opt. Commun., \textbf{219} (2003), 427-433.

\bibitem{Ricardo} R. Carretero-Gonz\'{a}les, J. D. Talley, C. Chong,  B.A. Malomed,
\emph{Multistable solitons in the cubic--quintic discrete nonlinear
Schr\"{o}dinger equation}, Physica D, \textbf{216} (2006), 77--89.

\bibitem{Ch09} C. Chong, R. Carretero-Gonz\'alez, B.A. Malomed, P.G. Kevrekidis,
\emph{Multistable solitons in higher--dimensional cubic--quintic
nonlinear Schr\"{o}dinger lattices}, Physica D, \textbf{238}
(2009), 126--136.

\bibitem{Demetri} D. N. Christodoulides, R. I. Joseph, \emph{Discrete self-focusing in nonlinear arrays of coupled waveguides}, Opt. Lett., \textbf{13} (1988), 794-796.

\bibitem{JKFM} J. Cuevas, P.G. Kevrekidis, D.J. Frantzeskakis,
B.A. Malomed, \emph{Discrete solitons in nonlinear Schr\"{o}dinger
lattices with a power-law nonlinearity}, Physica D, {\bf 238}
(2009), 67-76.

\bibitem{EJ03} J.Ch. Eilbeck, M. Johansson, \emph{The discrete nonlinear Schr{\"o}dinger equation—-20 years on}, in
``Localization and Energy Transfer in Nonlinear Systems"  (eds. L. Vazquez, R.S. MacKay, M.P. Zorzano), World Scientific, (2003), 44-67.

\bibitem{Stepic} L. Had\v{z}ievski, A. Maluckov, M. Stepi\'{c}, D. Kip, \emph{Power controlled soliton stability and steering in lattices with saturable nonlinearity}, Phys. Rev.
Lett., \textbf{93} (2004), 033901.

\bibitem{CQ3} R. A. Ganeev, M. Baba, M. Morita, A. I. Ryasnyansky, M. Suzuki,
M. Turu, H. Kuroda, \emph{Fifth-order optical nonlinearity of
pseudoisocyanine solution at 529 nm}, J. Opt. A: Pure Appl. Opt.
\textbf{6} (2004), 282-287.

\bibitem{Ka05} D.J.\ Kaup, \emph{Variational solutions for the discrete nonlinear Schrödinger equation},  Math. Comput. Simulat., \textbf{69} (2005), 322--333.

\bibitem{KRB01} P.G. Kevrekidis, K.{\O}. Rasmussen, A.R. Bishop, \emph{The discrete nonlinear Schr{\"o}dinger equation: a survey of recent results}, Int. J. Mod. Phys. B, \textbf{15} (2001), 2833-2900.

\bibitem{Loyd1} D.J.B. Lloyd, B. Sandstede, \emph{Localized radial solutions of
the Swift-Hohenberg equation}, Nonlinearity, {\bf 22} (2009), 485--524.

\bibitem{Ma02} B.A. Malomed, \emph{Variational methods in nonlinear fiber optics and related fields},
Prog. Opt., \textbf{43} (2002), 71-193.

\bibitem{MM91} B.A. Malomed, M.I. Weinstein, \emph{Soliton dynamics in the discrete nonlinear Schr\"odinger equation}, Phys. Lett. A, \textbf{220} (1996), 91-96.

\bibitem{Belgrade1} A. Maluckov, L. Had\v{z}ievski, B. A. Malomed, \emph{Dark solitons in dynamical lattices with the cubic-quintic nonlinearity}, Phys. Rev. E, \textbf{76} (2007), 046605.

\bibitem{Belgrade2} A.\ Maluckov, L. Had\v{z}ievski, B.A. Malomed, \emph{Staggered and moving localized modes in dynamical lattices with the cubic-quintic nonlinearity}, Phys. Rev. E, \textbf{77} (2008), 036604.

\bibitem{OJ} M. \"{O}ster, M. Johansson, \emph{Stability, mobility and power currents
in a two-dimensional model for waveguide arras with nonlinear
coupling}, Physica D, {\bf 238} (2009), 88--99.

\bibitem{OJE} M. \"{O}ster, M. Johansson, A. Eriksson, \emph{Enhanced mobility of strongly localized modes in waveguide arrays by inversion of stability Phys.}, Rev. E {\bf 67} (2003), 056606.

\bibitem{PKMF03} I.E. Papacharalampous, P.G. Kevrekidis, B.A. Malomed, D.J. Frantzeskakis,
\emph{Soliton collisions in the discrete nonlinear Schr\"odinger equation},
Phys. Rev. \textbf{68} (2003), 046604.

\bibitem{VJ} R.A. Vicencio, M. Johansson, \emph{Discrete soliton mobility in two-dimensional waveguide arrays with saturable nonlinearity}, Phys. Rev. E {\bf 73}
(2006), 046602.

\bibitem{CQ2} C. Zhan, D. Zhang, D. Zhu, D. Wang, Y. Li, D. Li, Z. Lu,
L. Zhao, Y. Nie, \emph{Third- and fifth-order optical nonlinearities in a new stilbazolium derivative}, J. Opt. Soc. Am. B,  \textbf{19} (2002), 369-375.

\end{thebibliography}
\end{document}